\documentclass{PoS}

\usepackage{subfig}

\title{Mass Composition of Cosmic Rays with Combined Surface Detector Arrays}

\ShortTitle{Mass Composition of Cosmic Rays with Combined Surface Detector Arrays}

\author{\speaker{Jakub V\'icha}$^a$, Dalibor Nosek$^{b}$, Petr Tr\'avn\'i\v{c}ek$^{a}$ and Jan Ebr$^{a}$\\
\llap{$^a$}Institute of Physics of the Czech Academy of Sciences, Prague, Czech Republic\\
Na Slovance 1999/2, 182 21 Praha 8, Czech Republic\\
\llap{$^b$}Charles University, Faculty of Mathematics and Physics, Prague, Czech Republic\\
V Hole\v{s}ovi\v{c}k\'{a}ch 2, 180 00 Praha 8, Czech Republic\\

E-mails: \email{vicha@fzu.cz}, \email{dalibor.nosek@mff.cuni.cz}, \email{petr.travnicek@fzu.cz}, \email{ebr@fzu.cz}}

\abstract{Our study exploits the Constant Intensity Cut principles applied simultaneously to muonic and electromagnetic detectors of cosmic rays. We use the fact that the ordering of events according to their signal sizes induced in different types of surface detectors provides information about the mass composition of primary cosmic-ray beam, with low sensitivity to details of hadronic interactions. Composition analysis at knee energies is performed using Monte Carlo simulations for extensive air showers having maxima located far away from a hypothetical observatory. Another type of a hypothetical observatory is adopted to examine composition of ultra-high energy primaries which initiate vertical air showers with maxima observed near surface detectors.}

\FullConference{35th International Cosmic Ray Conference --- ICRC 2017\\
		10--20 July, 2017\\
		Bexco, Busan, Korea}

\begin{document}

\section{Introduction}
The mass composition of cosmic rays is crucial to resolve their possible sources in the Universe. The fluorescence detection technique, see e.g. \cite{augerfd}, provides a very precise measurement of longitudinal profiles of cosmic-ray showers at ultra-high energies (above $\sim$10$^{18}$~eV) where the so-called ankle in the energy spectrum is observed \cite{augerspectrumicrc2015}. The mass composition is then usually derived from the measurement of the depth of shower maximum within these longitudinal profiles, see e.g. \cite{xmaxAuger}. The disadvantage of the fluorescence measurement is its low duty cycle ($\sim$10\%) due to requirements of moon-less and cloud-less nights. At energies where the so-called knees in the energy spectrum are observed ($10^{15-17}$~eV) \cite{Kascademass,KASCADE-Grande}, the fluorescence yield is so low that it is impossible to trigger and reconstruct cosmic-ray showers within this technique. On the other hand, the detection technique using huge arrays of particle detectors on the ground has almost full duty cycle above a full-efficiency threshold that depends on the detector spacings. However, a worse sensitivity to the mass of primary particles is obtained at ultra-high energies for an array of single type of surface detectors compared to the fluorescence measurements \cite{augerrisetime}.

The largest array ever-built to study cosmic-ray particles, the Pierre Auger Observatory \cite{augerNimA}, observes extensive showers of secondary particles in the atmosphere with a fully efficient array of surface detectors for energies above $10^{18.5}$~eV and zenith angles up to 60$^{\circ}$. The signals in water-Cherenkov detectors are induced by electromagnetic (EM) and muonic component of shower as well. The Observatory is located at altitude 1400~m a.s.l. which corresponds to $\sim$880~g/cm$^{2}$ of vertical atmospheric depth. At ultra-high energies, this depth is comparable with depths of shower maxima. The water-Cherenkov detectors are going to be upgraded with scintillator detectors in the next years \cite{augerupgrade} allowing to measure the EM component of showers separatelly in order to disentagle the sizes of EM and muonic components of showers to improve the sensitivity to the mass composition of cosmic rays.

The KASCADE experiment \cite{KASCADE} was equipped with coincident arrays of scintillator detectors measuring separatelly EM and muonic particles (shielded with iron/lead plates) of showers with full efficiency in the energy range $10^{15-17}$~eV and zenith angles below 40$^{\circ}$. KASCADE was located near the sea level (110 m a.s.l.) with vertical atmospheric depth $\sim$1020~g/cm$^{2}$ that is a much higher depth than the depths of shower maxima in the observed energy range. Combining the measurement of the number of electrons and of the number of muons, different groups of primary particles were selected to study cosmic-ray spectra \cite{Kascademass}.

Generally, the signal induced in the surface detectors is attenuated with zenith angle due to a different amount of atmosphere that is passed before the shower reaches the surface detector array. The EM and muonic signals attenuate differently mainly due to different energy losses in the air for EM particles and muons. Also, a different attenuation of signals in the surface detectors is caused by different primary particles mainly due to different distances of shower maximum to the ground. A data-driven method, the so-called Constant Intensity Cut (CIC) method \cite{hersil}, selects the highest signals at various zenith angles under the assumption of isotropic arrivals of cosmic rays at given intensity. The CIC method finds reliably the average attenuation of signal in the surface detectors for a composition at given flux without any assumption on the mass composition \cite{vichaicrc2013}.

In this study, we consider a hypothetical array of coincident surface detectors sensitive to the different components of shower. This hypothetical observatory is similar to the KASCADE experiment (denoted as KNEE Observatory in the following). We compare our method with our previous work published in \cite{vichaAstro} that considered a hypothetical observatory inspired by the surface detectors of the Pierre Auger Observatory upgraded with some type of muon detectors (ANKLE Observatory). Applying the CIC approach, we show a sensitivity to the mass composition of cosmic rays that has a weak dependence on details of hadronic interactions.

\section{Simulated Signals in Surface Detectors}
To exploit the details of the CIC method caused by different primaries a large simulated data sample, about $\sim$10$^{5-6}$ simulated showers, is necessary. This is comparable with the statistics of both of cosmic-ray experiments considered in this proceedings. Such excessive computational requirements were avoided generating sets of reference showers induced by proton (p), and helium (He), nitrogen (N) and iron (Fe) nuclei with energies and zenith angles ($\theta$) as indicated in Tab.~\ref{TabCorsikaSetting}. The zenith angles of incoming primaries were chosen to maintain equal steps in cos$^{2}(\theta)$. These showers were produced by CORSIKA ver.~7.37 \cite{corsika} with two models of high-energy hadronic interactions tuned to the LHC data (run I): QGSJet~II-04 \cite{qgsjet} and EPOS-LHC \cite{epos}. The hadronic interactions at lower energies (below 80~GeV) were treated with the FLUKA model \cite{fluka}. 

\begin{table}[!h]
\begin{center}
\caption{Parameters of reference CORSIKA showers. For each energy ($E_{\rm MC}$), zenith angle ($\theta$) and model of hadronic interactions a given number of showers ($N_{\rm Corsika}$) was produced for the two hypothetical observatories at different observation levels for each of $N_{\rm steps}$ steps in cos$^{2}(\theta)$ of $N_{\rm steps}$.}
\begin{tabular}{|c||c|c|c|c|c|}
\hline
Observatory & Altitude [m] & log($E_{\rm MC}$ [eV]) & $\theta [^{\circ}]$ & $N_{\rm steps}$ & $N_{\rm Corsika}$ \\
\hline\hline
KNEE & 110 & 15, 15.5, 16, 16.5, 17 & 0-40 & 10 & 100 \\
\hline
ANKLE & 1400 & 19 & 0-45 & 7 & 60 \\
\hline
\end{tabular}
\label{TabCorsikaSetting}
\end{center}
\end{table}

\begin{figure}[h!]
  \centering
\includegraphics[width=0.95\textwidth]{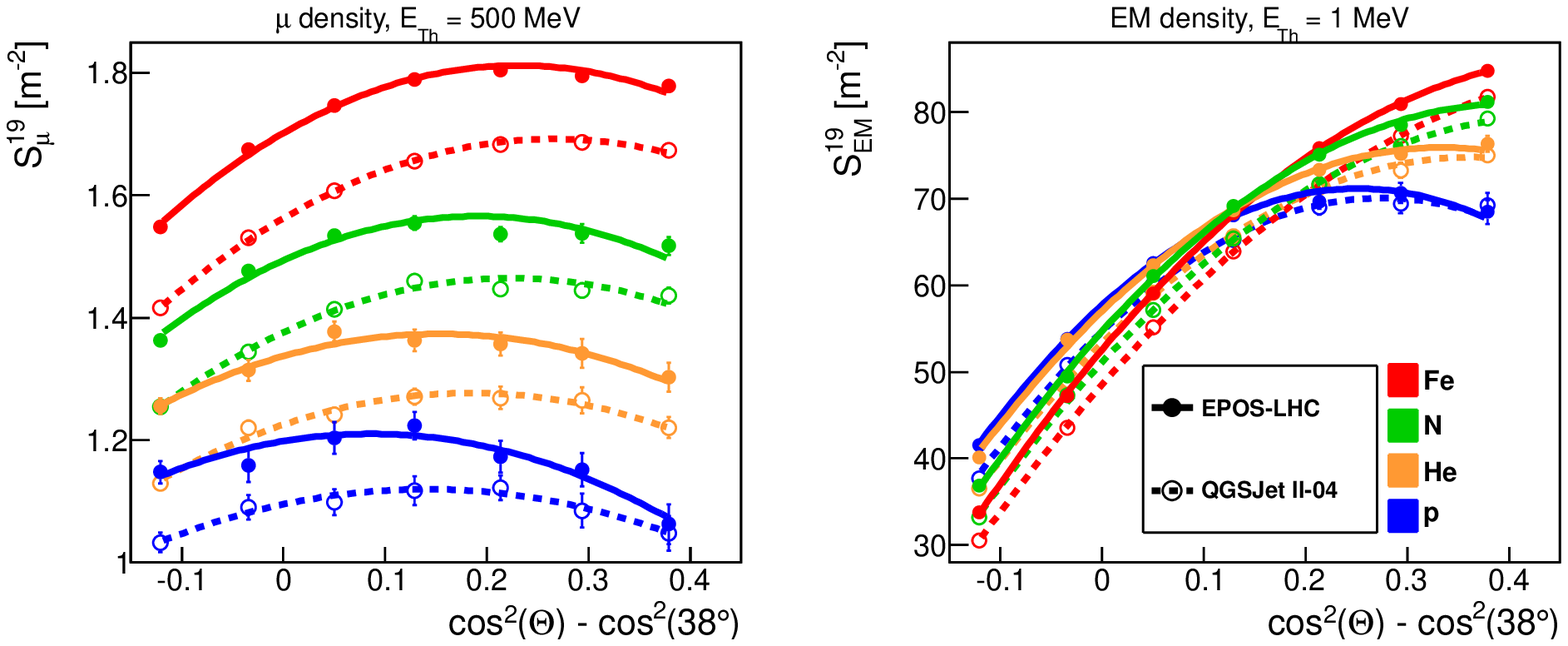}
\includegraphics[width=0.95\textwidth]{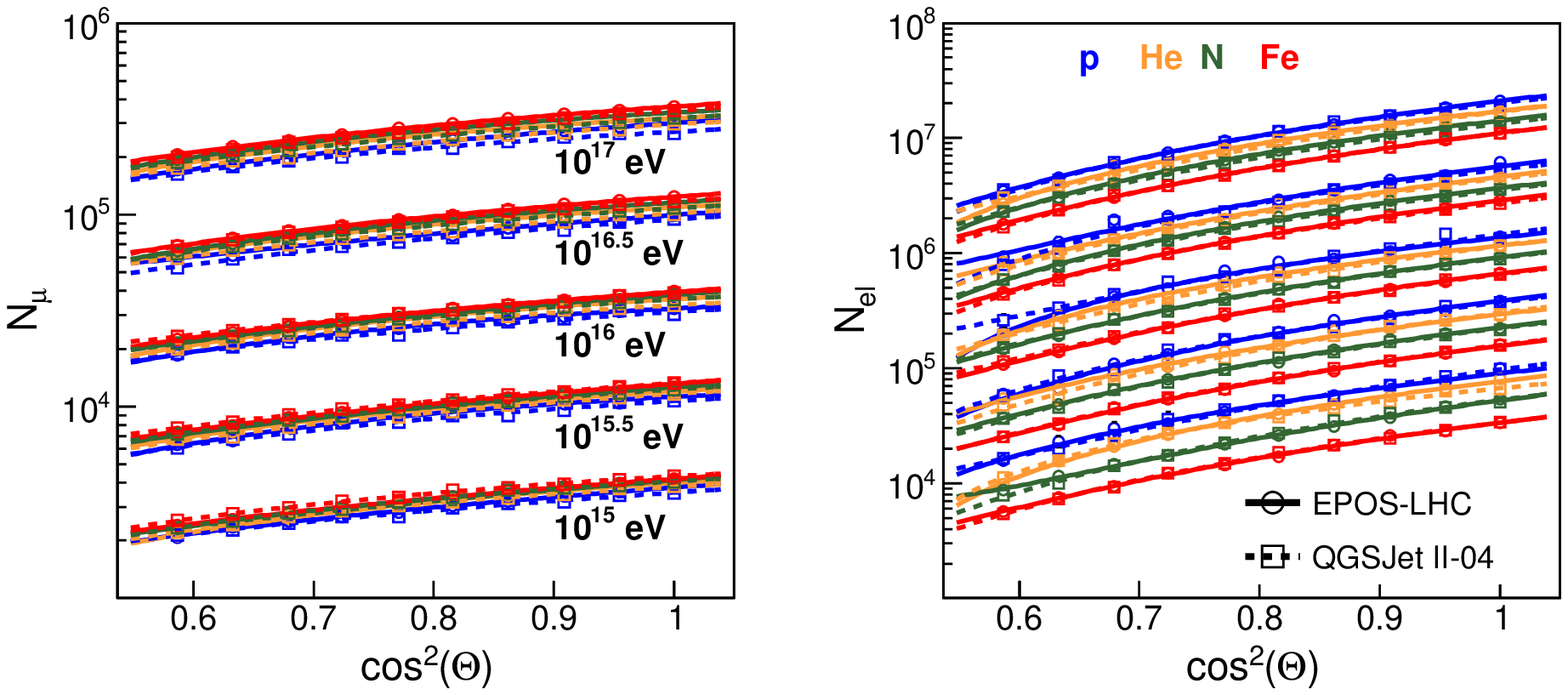}
\caption{Attenuation curves for the KNEE Observatory (bottom) and for the ANKLE Observatory (top, from \cite{vichaAstro}). Reference signals of CORSIKA showers are fitted with quadratic functions of $\cos^{2}(\Theta)$ for the muon detectors (left panels) and for the EM detectors (right panels). Two models of hadronic interactions and four primary species are distinguished by types of lines and colors, respectively.}
  \label{CICfit}
\end{figure}

The reference signals (Fig.~\ref{CICfit}) induced in the two arrays were derived according to the threshold energies and shower core distances of muons and EM particles as indicated in Tab.~\ref{TabParticleSampling}. In the case of the KNEE Observatory, we considered that the detector responses are proportional to the number of muons ($N_{\mu}$) and electrons ($N_{\rm el}$) (shielded and unshielded scintillators). In the case of the ANKLE Observatory, we assumed that the detector responses are proportional to the muon density and EM density which was motivated by the responses of water-Cherenkov detectors.

\begin{table}[!h]
\begin{center}
\caption{Cuts on secondary particles. EM and muonic particles with threshold energies $E_{\rm th}(\rm EM)$ and $E_{\rm th}(\mu)$ were sampled in circular regions with radius $r_{\rm EM}$ and $r_{\mu}$ perpendicular to the shower axis, respectively. In the case of the ANKLE Observatory, a fit to the lateral distribution function was performed to obtain signals at 1000~m from the shower core.}
\begin{tabular}{|c||c|c|c|c|}
\hline
Observatory & $E_{\rm th}(\rm EM)$ [MeV] & $E_{\rm th}(\mu)$ [MeV] & $r_{\rm EM}$ [m] & $r_{\mu}$ [m] \\
\hline\hline
KNEE & 5 & 230 / cos($\theta$) & $<0,\infty>$ & $<40,200>$ \\
\hline
ANKLE & 1 & 500 & 1000 & 1000 \\
\hline
\end{tabular}
\label{TabParticleSampling}
\end{center}
\end{table}

\begin{table}[!h]
\begin{center}
\caption{Parameters to generate signals in continuous range of energies ($E_{\rm Gen}$) and zenith angles ($\theta_{\rm Gen}$). The generated energy was distributed according to the energy spectrum $\propto E_{\rm Gen}^{-\gamma}$ with the total number of events $N_{\rm Gen}$ for each model and each mass composition of primaries. For the ultra-high energies, a steep decrease of the cosmic-ray flux was considered according to \cite{augerspectrumicrc2013}. The muonic and EM signals were additonally smeared by Gaussian with relative width $\sigma_{\mu}$ and $\sigma_{EM}$, respectively.}
\begin{tabular}{|c||c|c|c|c|c|c|}
\hline
Observatory & log( $E_{\rm Gen}$ [eV] ) & $\gamma$ & $\theta_{\rm Gen}$ [$^{\circ}$] & $\sigma_{\mu}$ [\%] & $\sigma_{EM}$ [\%] & $N_{\rm Gen}$\\
\hline\hline
KNEE & 15-17 & 3.0 & 0-40 & 30-7 & 40-3& $10^{5}$ \\
\hline
ANKLE & 18.5-20 & 2.7$\rightarrow\sim$6 & 0-45 & 20 & 20 & $7\cdot10^{5}$ \\
\hline
\end{tabular}
\label{TabSignalProduction}
\end{center}
\end{table}

These reference signals were parametrized as functions of energy and zenith angle together with the respective fluctuations and correlations of muonic and EM signals. These parametrized curves (polynoms up to the second order) were utilized in the simplified simulation of the muonic and EM signals induced by showers over a wide range of energies and zenith angles (see Tab.~\ref{TabSignalProduction}). Additional smearing was applied to the generated signals to account for the detector resolutions according to \cite{KascadeResolution} (improving with energy) in the case of the KNEE Observatory and by some realistic assumptions at ultra-high energies for the ANKLE Observatory.

\begin{figure}[h!]
  \centering
\includegraphics[width=1.0\textwidth]{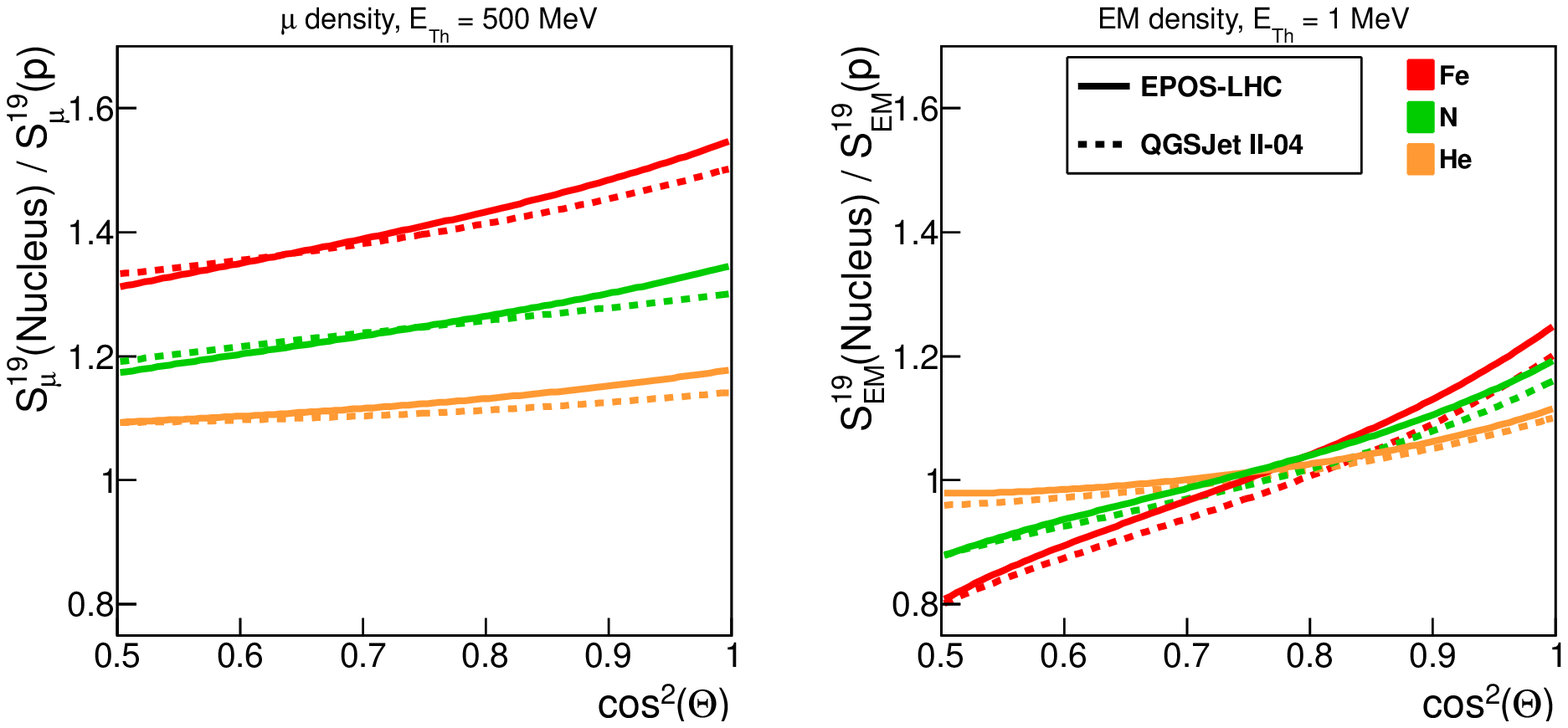}
\includegraphics[width=0.5\textwidth]{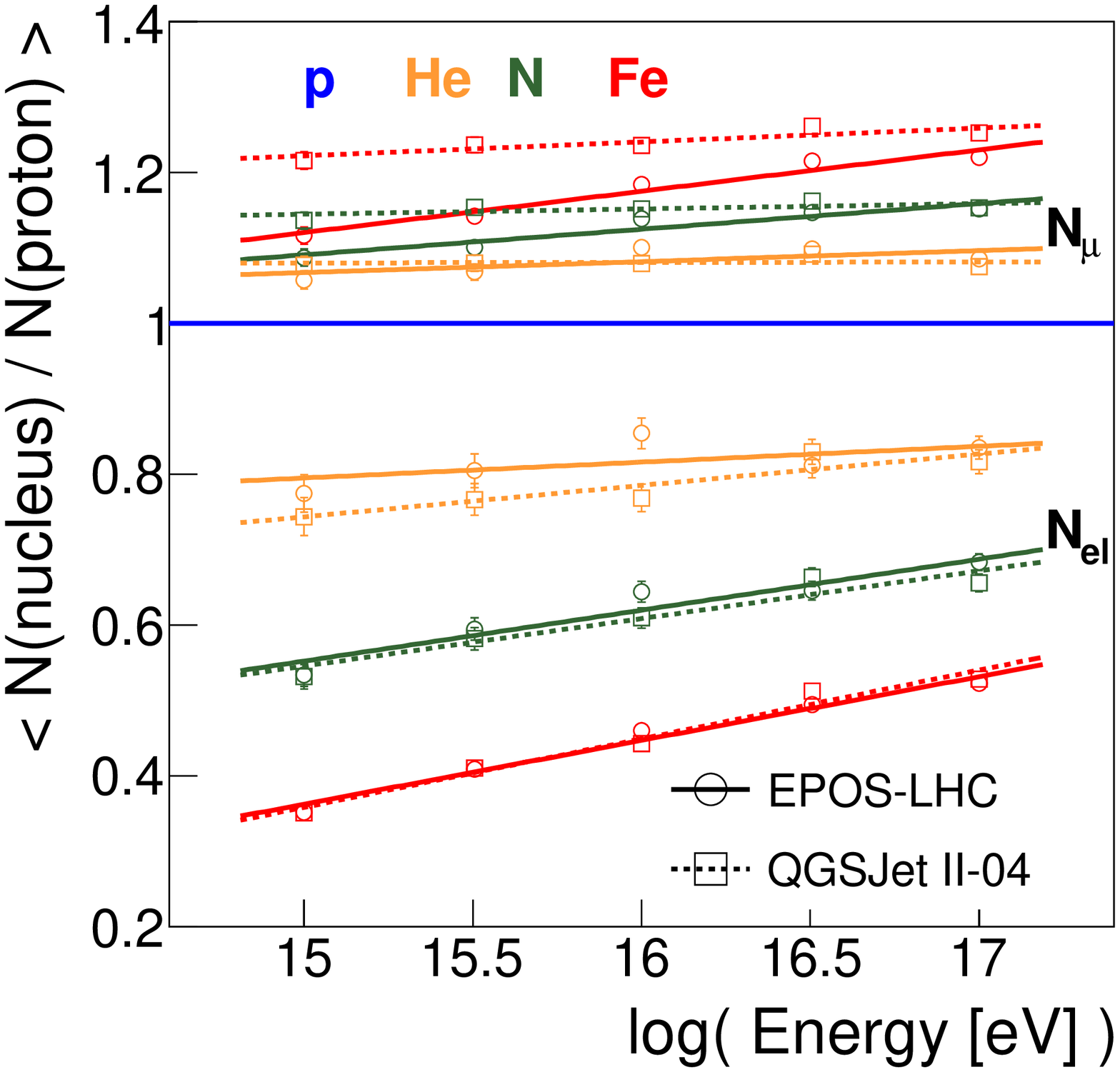}
  \caption{Ratios of shower signals for the KNEE Observatory (bottom) and for the ANKLE Observatory (top, from \cite{vichaAstro}). Ratios for the muon (top left panel) and EM (top right panel) signals induced by nuclei and protons are plotted as a function of cos$^{2}(\Theta)$. Ratios for the number of muons (greater than 1) and for the number of electrons (smaller than 1) in showers generated by nuclei and protons are plotted as a function of energy on the bottom panel. Two models of hadronic interactions and primary nuclei are distinguished by types of lines and colors, respectively.}
  \label{NucleiToProtonRatios}
 \end{figure}

\section{Observatory close to Shower Maximum}
In the case of cosmic-ray showers at ultra-high energies that are observed at 880~g/cm$^{2}$ of vertical atmospheric depth, the depths of shower maximum are located very close to the observation level and therefore a significant difference in the relative attenuation of signals induced by different primaries is observed for both signals, see the top panels of Fig.~\ref{CICfit}. This has a consequence that the ratio of signals induced by different primaries has a strong dependence on zenith angle in the case of EM signal and a moderate trend in the case of muonic signal (see top panels of Fig.~\ref{NucleiToProtonRatios}). Note that the ratio for EM signal becomes even smaller than 1 for higher zenith angles. 

\begin{figure}[h!]
  \centering
	\subfloat{\includegraphics[width=0.53\textwidth]{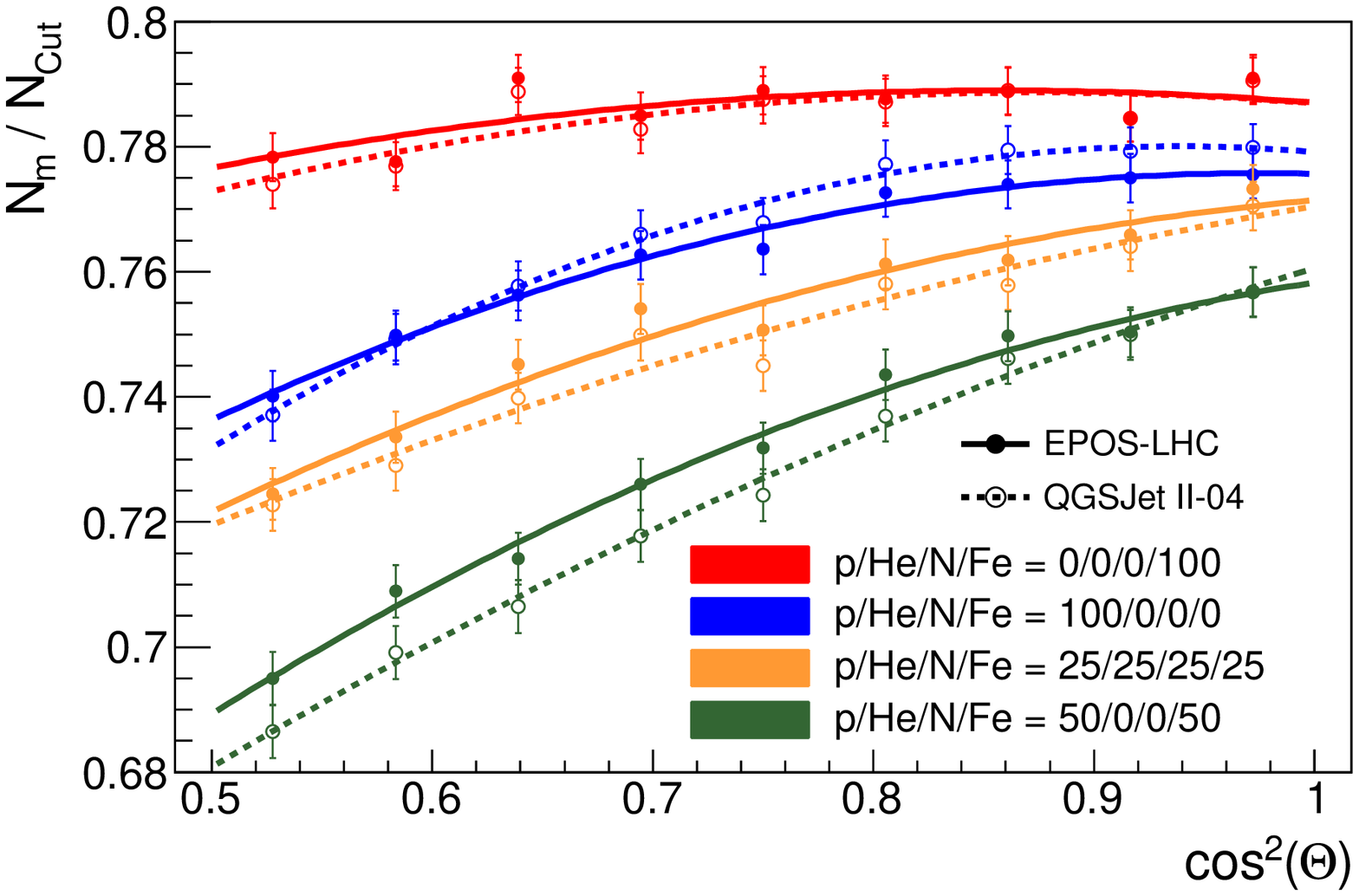}}
	\subfloat{\includegraphics[width=0.47\textwidth]{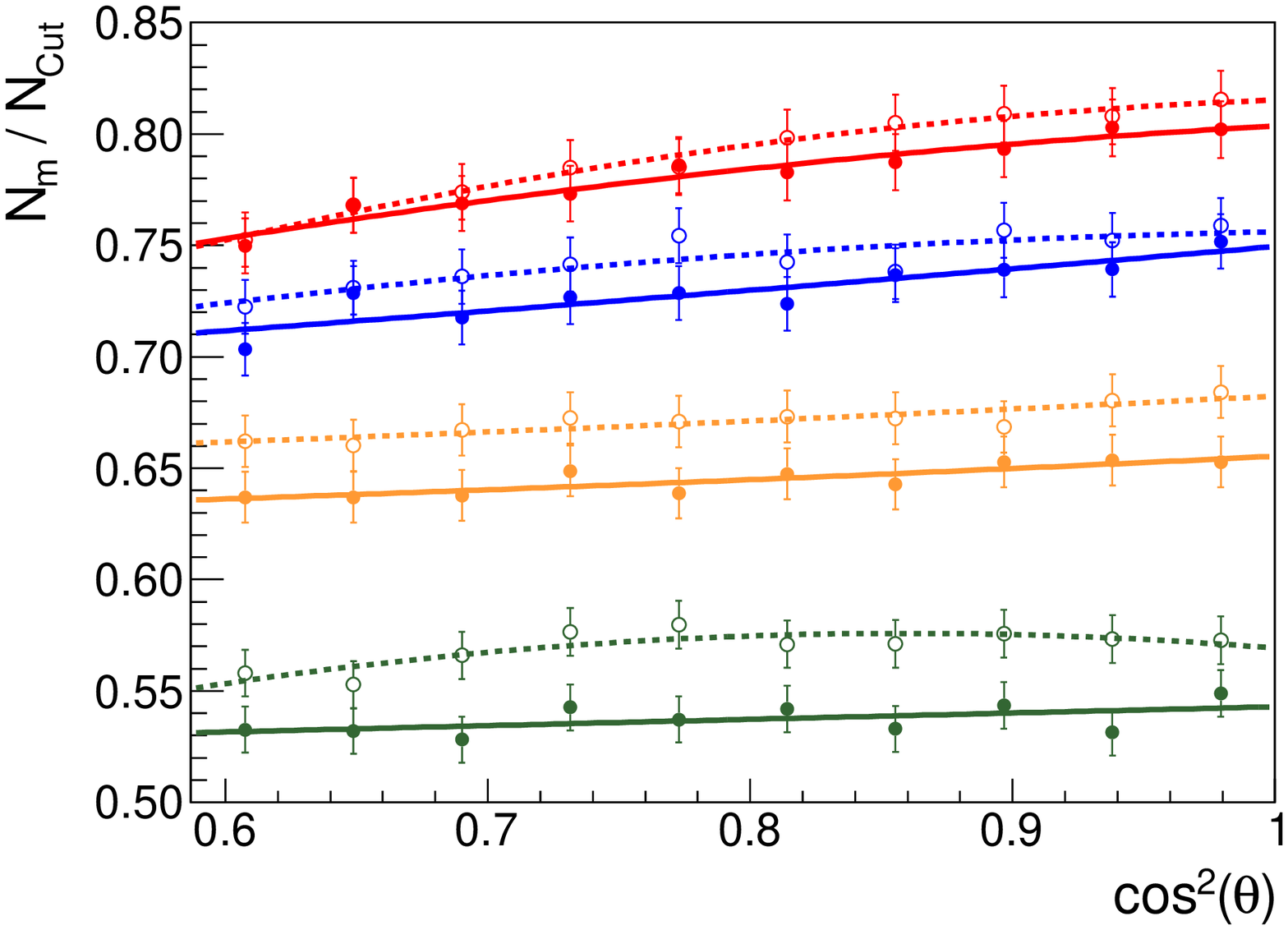}}
  \caption{Attenuation of the matched fraction for the ANKLE Observatory (left, from \cite{vichaAstro}) and for the KNEE Observatory (right). Fractions of simulated events that are matched in both detectors, $N_{\rm m}/N_{\rm Cut}$, are plotted as a function of $\cos^{2}(\theta)$. Results of four examples of the primary mass composition (different colors) and for the two models of hadronic interactions (full and empty markers) are shown. Depicted curves are quadratic fits to these fractions.}
  \label{MatchedNumber}
 \end{figure}

This feature was utilized in our previous work \cite{vichaAstro} calculating the number of events that are matched ($N_{\rm m} = \rm| \mathbb{M}^{\mu} \cap \mathbb{M}^{\rm EM} |$ $\leq N_{\rm Cut}$) in the two sets $\mathbb{M}^{\mu}$ and $\mathbb{M}^{\rm EM}$ each of $N_{\rm Cut}$ events with the largest values (the CIC principle) of muonic and EM signals, respectively. The more the primary beam is mixed, the higher difference in the ordering of the same events (smaller $N_{\rm m}$) is obtained in the muon and EM signals. The relative change of $N_{\rm m}/N_{\rm Cut}$ with zenith angle was found to be sensitive to the spread of primary masses, see left panel of Fig.~\ref{MatchedNumber} for $N_{\rm Cut}=12000$ out of $7\times10^{5}$ events. As an example, we show results for single protons (blue), single Fe nuclei (red), a mix of protons and He, N, Fe nuclei (orange), and a mix of protons and Fe nuclei with the maximal variance of logarithmic mass number of the considered primaries (green). The absolute value of $N_{\rm m}/N_{\rm Cut}$ is actually sensitive to the spread of primary masses as well, but the relative change with zenith angle is less dependent on the model of hadronic interactions and does not depend on the detector resolutions that much. This was demonstrated with parameter $\Phi=1-<N_{\rm m}/N_{\rm Cut}>(45^{\circ})/<N_{\rm m}/N_{\rm Cut}>(0^{\circ})$ for the two models of hadronic interactions and 3 ranges of mean logarithmic mass. The values of ${<N_{\rm m}/N_{\rm Cut}>(\theta)}$ were obtained from quadratic fits as indicated on the left panel of Fig.~\ref{MatchedNumber}. The results were very similar for the two models of hadronic interactions and the variance of logarithmic mass number could be achieved with similar uncertainty as in the case of fluorescence measurements.

\section{Observatory far from Shower Maximum}
In the case of the KNEE Observatory, the vertical atmospheric depth was considered to be 1020 g/cm$^{2}$ and the depths of shower maximum occur very far from the surface detectors; especially at the energies $10^{15-17}$~eV. This implies that the differencies between the shapes of attenuation curves for different primaries are very small (see bottom panels of Fig.~\ref{CICfit}). Consequently, the ratios of signals induced by different nuclei are approximately constant with zenith angle. 

On the bottom panel of Fig.~\ref{NucleiToProtonRatios}, the evolution of the signal ratios with energy is shown. The proton-induced $N_{\rm el}$ is always greater than nuclei-induced $N_{\rm el}$ in the considered energy range; opposite in the case of $N_{\mu}$. However, extrapolating $N_{\rm el}$ to the energy $10^{19}$~eV and considering the difference in the vertical atmospheric depth (140~g/cm$^{2}$) and the difference in the shower core distance ($\geq$~800~m) between the two observatories, we would obtain comparable EM signals for different primaries which is what was observed and exploited in the case of the ANKLE Observatory.

The parameter $\Phi$, as defined for the ANKLE Observatory, varies within few \% for all the combinations of mass composition of the four primaries and it would be therefore inefficient to derive the mass composition of cosmic rays at the KNEE Observatory with this parameter. Instead of $\Phi$, we propose to study e.g. $\psi=<N_{\rm m}/N_{\rm Cut}>(0^{\circ})$ at the KNEE Observatory. On the right panel of Fig.~\ref{MatchedNumber}, the sensitivity to the spread of primary masses is indicated for $N_{\rm Cut}=5000$ out of $10^{5}$ events. The parameter $\psi$ depends more on the knowledge of detector resolutions and shower fluctuations than in the case of parameter $\Phi$. However, the KASCADE Collaboration determined their resolutions for both types of detectors in \cite{KascadeResolution}, and, moreover, the shower fluctuations differs much less for different models of hadronic interactions at energies $10^{15-17}$~eV than for ultra-high energies. Therefore we intend to apply this method to KASCADE data trying to estimate the variance of logarithmic mass number as a function of energy in our next work.

\section{Conclusions}
We have produced sets of CORSIKA showers for different models of hadronic interactions, primaries and zenith angles. The resulting reference signals that would be induced in a hypothetical arrays of muonic and EM detectors (inspired by the KASCADE experiment) were parametrized to generate a large sets of signals over a wide range of energies and zenith angles. With these sets of signals, we have indicated a sensitivity of the fraction of events matched in both detectors (within the CIC approach) to the mass composition of primary cosmic rays. The analysis was compared with our previous work considering a hypothetical observatory at ultra-high energies. The presented method seems to be advantageous due to low sensitivity to details of hadronic interactions to complement frequently conducted studies of the mass composition that are based on the analysis of the mean logarithmic mass of primary species.

\section*{Acknowledgements}
This work is funded by Ministry of Education, Youth and Sports of the Czech Republic under the projects LG 15014 and LM 2015038. This work was also supported by ESIF and MEYS (Project AUGER.CZ - CZ.02.1.01/0.0/0.0/16\_013/0001402) and by the Czech Science Foundation under project 14-17501S.

\end{document}